\documentclass[aps,prb,twocolumn,showpacs]{revtex4}

\usepackage{graphicx}

\usepackage{times}

\begin{document}

\title{Effects of phonon interaction on the pairing in the high-T$c$ superconductors}

\author{Yunkyu Bang}
\affiliation{Department of Physics, Chonnam National University,
Kwangju 500-757, and Asia Pacific Center for Theoretical Physics, Pohang 790-784, Korea}

\begin{abstract}
We study the effects of phonon interaction on the superconducting
pairing in the background of a d-wave gap, mediated by
antiferromagnetic (AFM) spin fluctuations, using coupled BCS gap
equations. We found that phonon interaction can induce a s-wave
component to the d-wave gap in the (D+S) form with an interaction
anisotropy and in the (D+$i$S) form without anisotropy,
respectively. In either case, however, T$_c$ is not enhanced
compared to the pure d-wave pairing without phonon interaction. On
the other hand, anisotropic phonon interaction can dramatically
enhance the d-wave pairing itself and therefore T$_c$, together
with the AFM spin fluctuation interaction. This (D$_{AFM}$ +
D$_{ph}$) type pairing exhibits strongly reduced isotope
coefficient despite the large enhancement of T$_c$ by phonon
interaction. Finally, we study the combined type of (D$_{AFM}$ +
D$_{ph}$ +$i$S)) gap and calculate the penetration depth and
specific heat to be compared with the experiments.

\end{abstract}

\pacs{74.20,74.20-z,74.50}

\date{\today}
\maketitle

\section{Introduction}

The problems of the high T$_c$ superconductors (HTSC) have been
mostly focused on the electronic correlations and phonons were
usually considered as a secondary player at best. However, recent
Angle Resolved Photoemission Spectroscopy (ARPES) experiments
revived the interest of the possibly important role of phonons in
the high-T$_c$ cuprates (HTC)\cite{review 06}.
In particular, the systematic measurements and analysis \cite{cuk
04} of the kink structures in the quasiparticle dispersion near
the Fermi surface (FS) proved that: (1) the origin of the kink is
electron-phonon coupling and its coupling is very strong, (2) the
typical energy of the phonon(s) is $\sim$ 40 - 70 meV, and (3) the
coupling matrix is quite anisotropic. Therefore, it is currently a
pressing question what possible roles and effects, in particular,
for the superconducting (SC) pairing, the phonon(s) can do in the
HTC.

Although the SC gap symmetry in the high-T$_c$ cuprates is well
established as a d-wave by most experiments, there are also
continuous experimental reports such as tunnelling conductance
\cite{tunnelling}, penetration depth measurements\cite{penet}, etc
that provide rather convincing evidences for a small magnitude of
a s-wave component in addition to the dominant d-wave SC gap in
some of HTC compounds. And it is well known that phonons mediate
an attractive interaction between electrons and lead to an
isotropic s-wave gap. Therefore, more sharpened question will be:
is it possible to reconcile a d-wave gap and a s-wave gap together
? and in that case, what is the role of phonon(s) ?

While there is no consensus yet on the pairing mechanism for the
d-wave gap in HTC, in this paper we assume that the AFM spin
fluctuations is the mediating glue for d-wave gap \cite{Pines}.
The key ingredient of this mechanism is that the sign changing
character of the d-wave gap turns the all positive definite
potential of the AFM spin fluctuations mediated interaction (we
will call it "AFM interaction" for short from now), in momentum
space and in spin singlet channel, into an effectively attractive
pairing interaction.
From this point of view, it is clear that the phonon interaction
and AFM interaction are antagonistic and difficult to work
together for making a pairing because they tend to promote a
different symmetry of the SC gap each other.

An easy way to reconcile the d-wave and s-wave gap with two
different pairing interactions would be to invoke two separate
bands for each gap\cite{coupled gap}. The similar idea for the two
s-wave gaps was very successful for MgB$_2$ \cite{two gap} because
there two well separated bands exist. However, in HTSC, numerous
experiments, in particular ARPES measurements\cite{Shen RMP}, show
that there is only one main band crossing FS.
Then any idea of having s-wave and d-wave gaps in the single band,
induced by two very different pairing interactions, seems too
naive. However, if the coupling matrix of phonon interaction
possesses a strong anisotropy, this combined pairing problem is
not a trivial one and needs systematic investigation with a
traceable formulation. This problem has been already studied by
many authors with different techniques and
aims\cite{polaron,honer,Zeyher}. In this paper, we took a simplest
approach and studied the coupled gap equations for multiple gaps
of a single band with two pairing interactions, i.e., AFM
interaction and anisotropic phonon interaction, within the BCS
framework. More justification for this approach will be discussed
in the next section. By numerical solutions, we extensively
investigated the necessary conditions of the strength and the
degree of anisotropy of  phonon interaction for various multigap
solutions.

We found that the (D+S) and (D+$i$S) type solutions are indeed
possible with a proper degree of anisotropy of phonon interaction
for each case. However, T$_c$ is not enhanced at all in these
cases even though an additional s-wave component is formed due to
the phonon interaction. Other interesting possibility is that the
anisotropic phonon interaction can induce an additional d-wave
component ($D_{ph}$) to the AFM interaction induced d-wave
component ($D_{AFM}$). As a result, T$_c$ is dramatically enhanced
due to the phonon interaction.
We then derived an analytic T$_c$ equation for this
($D_{AFM}+D_{ph}$) case and showed that phonon isotope effect can
be strongly reduced due to the interplay between the AFM and
phonon interactions despite the large increase of T$_c$ by phonon.
This result can explain a long standing puzzle of the small phonon
isotope effect in HTSC. In view of experiments\cite{cuk 04}, the
best anisotropic phonon can be the B$_{1g}$ buckling phonon mode
which fulfills all necessary qualifications for our model. Similar
conclusion was obtained by other authors\cite{honer} using
different approaches. We emphasize that the final effect of
phonons in HTC compounds should be judged by an average effect of
all important phonons together, which is beyond the scope of this
paper.

Finally, we also considered the combined type of
($D_{AFM}+D_{ph}+i$S) gap which is a very natural solution of the
coupled gap equations and shows features such as (1) large
enhanced T$_c$, and (2) appearance of a s-wave component at low
temperatures. This type of gap will be a possible solution for the
tunnelling conductance experiment in YBCO\cite{tunnelling}.
Calculations of superfluidity density also demonstrated that
($D_{AFM}+D_{ph}+i$S) gap can explain the non-monotonic
temperature dependence of the penetration depth measurements in
YBCO and LSCO compounds \cite{penet}.

\section{Formalism}

The interplay between phonons and electronic correlations in HTSC
has been studied by numerous authors using different theoretical
approaches and models \cite{polaron,honer,Zeyher}. In this paper,
we took a simple minded approach with a specific question for this
problem. Experiments tell us that, after all correlation effects
taken place, the quasiparticles (q.p.) remain to form a fermi
surface (FS)\cite{Shen RMP}. And quite detailed properties of the
AFM spin fluctuations and symmetry allowed phonons are
experimentally measured\cite{review 06}; theses measured bosonic
fluctuations are final outcomes after all interplays between them
and correlation effects. Now on the phenomenological basis, we
consider a q.p. band and the AFM spin fluctuations as a dominant
pairing interaction. In addition to that we add anisotropic phonon
interactions and we study the pairing instability of the model for
several possible types of multi-gap solution using a generalized
BCS gap equations. In this approach, still remaining interplay
between spin fluctuations and phonons is ignored. The results of
this paper should be taken with this caveat. The Hamiltonian is
written as

\begin{eqnarray}
H &=& \sum_{k \sigma} \epsilon(k) c^{\dag}_{k \sigma} c_{k \sigma}
+ \sum_{k k^{'} \uparrow \downarrow} V_{AFM}(k,k^{'}) c^{\dag}_{k
\uparrow} c^{\dag}_{-k \downarrow}
c_{k^{'} \downarrow}c_{-k^{'} \uparrow} \nonumber \\
& & \sum_{k k^{'} \uparrow \downarrow} V_{ph} (k,k^{'})
c^{\dag}_{k \uparrow} c^{\dag}_{-k \downarrow} c_{k^{'}
\downarrow}c_{-k^{'} \uparrow}
\end{eqnarray}

\noindent where $\epsilon(k)$ is the dispersion of the
quasiparticles created by $c^{\dag}_{k \sigma}$ as standard
notation. $V_{AFM}(k,k^{'})$ and $V_{ph}(k,k^{'})$ are the
effective interactions, for the singlet superconducting pairing
channel, originating from the AFM spin fluctuations and phonon(s),
respectively. For traceable numerical calculations, we further
simplify the above Hamiltonian as follows. The real two
dimensional FS is simplified as a circular FS and the interactions
are also modelled accordingly as follows.
\begin{equation}
V_{AFM} (\Delta \phi) = V_M \frac{\phi_0 ^2}{(\Delta \phi \pm
\phi_{AFM})^2 +\phi_0 ^2}
\end{equation}
and
\begin{equation}
V_{ph} (\Delta \phi) = \left\{ \begin{array} {ll} - V_P &
\textrm{for~~} |\Delta \phi| < \phi_{AN} \\
0 & \textrm{for~~} |\Delta \phi| > \phi_{AN}
\end{array} \right.
\end{equation}

\noindent where $\Delta \phi= \phi -\phi^{'}$ and
$\phi_{AFM}=\pi/2$ representing the exchanged momentum ${\bf k -
k^{'}}$ and the AFM ordering vector ${\bf Q}$ in the circular FS,
respectively. $V_{M}$ and $V_{P}$ are chosen to be positive, so
that $V_{AFM}$ is all repulsive and $V_{ph}$ is all attractive in
momentum space.

The property of the AFM interaction with a short range correlation
is simulated with the inverse correlation length parameter
$\phi_0$  ($\lambda=(\pi a / \sqrt{2}) \phi_0$). In this paper, we
chose $\phi_0$=1 ($\lambda \sim 2 a$) for all numerical
calculations, which is quite a short range AFM correlation. The
degree of anisotropy of the phonon interaction is controlled by
the anisotropy angle parameter $\phi_{AN}$ which restricts the
scattering angle between incoming and outgoing q.p momenta; for
example, $\phi_{AN}= \pi$ would allow a perfectly isotropic
interaction. The phonon interaction $V_{ph} (\phi -\phi^{'})$,
however, does not restrict the incoming ($\phi$) and outgoing
momenta ($\phi^{'}$). Now the reduced BCS Hamiltonian in the mean
field theory can be written as

\begin{eqnarray}
H &=& \sum_{\phi \xi \sigma} \epsilon(\xi) c^{\dag}_{\phi \xi
\sigma} c_{\phi \xi \sigma} + \sum_{\phi \xi} \Delta_{AFM} ^{*}
(\phi) c_{\phi \xi \downarrow} c_{-\phi \xi \uparrow} \nonumber
\\ & &+ \sum_{\phi \xi} \Delta_{ph} ^{*} (\phi) c_{\phi \xi \downarrow}
c_{-\phi \xi \uparrow}
\end{eqnarray}

\noindent where $\Delta_{AFM}(\phi)$ is the SC gap function
induced by $V_{AFM}$ and $\Delta_{ph}(\phi)$ is the one induced by
$V_{ph}$. The two gap functions $\Delta_{AFM}(\phi)$ and
$\Delta_{ph}(\phi) $ may or may not have the same symmetry. After
diagonalizing the above Hamiltonian we obtain two self-consistent
equations as

\begin{eqnarray}
\Delta_{AFM} (\phi) & = &  \sum_{\phi^{'} \xi} V_{AFM}(\phi -
\phi^{'}) <c_{\phi \xi \downarrow} c_{-\phi \xi \uparrow}> , \\
\Delta_{ph} (\phi) & = &  \sum_{\phi^{'} \xi} V_{ph}(\phi -
\phi^{'}) <c_{\phi \xi \downarrow} c_{-\phi \xi \uparrow}>.
\end{eqnarray}

Similar coupled gap equations were studied in previous studies
\cite{coupled gap}. The key difference is that our model has only
one band. Then because the same band electrons should form the
multigaps, there is severe competition between different gaps and
more constraint to allow multigap solutions. In the next sections,
we will consider several possible multigap solutions of the above
coupled gap equations.

\subsection {(D+S) case}

In this case, we assume $\Delta_{AFM}(\phi)=\Delta_{d} \cos(2
\phi)$ and $\Delta_{ph}(\phi)=\Delta_{s}$. This leads to the two
coupled gap equations as follows.
\begin{eqnarray}
\Delta_d (\phi) & = & - \sum_{\phi^{'}} V_{AFM}(\phi - \phi^{'})
\Delta_t (\phi^{'})  \chi(\phi^{'},\omega_{AFM}), \\
\Delta_s (\phi) & = &  -\sum_{\phi^{'}} V_{ph}(\phi - \phi^{'})
\Delta_t (\phi^{'}) \chi(\phi^{'},\omega_p), \\
\chi(\phi^{'},\omega_{AFM,ph}) &=& N(0) \int _0 ^{\omega_{AFM,ph}}
d \xi \frac{\tanh (\frac{E(\phi^{'})}{2 T})}{E(\phi^{'})}
\end{eqnarray}

\noindent where $E(\phi) = \sqrt {\xi^2 + \Delta_{t}^2(\phi)}$,
and $\Delta_{t} (\phi)=(\Delta_{d} \cos(2 \phi)+\Delta_{s}$) is
the total gap function. $N(0)$ is the density of states (DOS) at
FS and $\omega_{AFM,ph}$ are the BCS energy cutoffs of each
pairing interactions $V_{AFM}$ and $V_{ph}$, respectively. Nonzero
value of $\Delta_{s}$ reduces the symmetry S$_4$ of the pure
d-wave gap to C$_2$; namely, the nodal points shift away from
diagonal directions and the sizes of the positive lobe and
negative lobe of the total gap function $\Delta_{t} (\phi)$ become
different.
We found that the coexistence of d- and s-wave gaps is possible
only when $V_{ph}(\phi-\phi^{'})$ is strongly anisotropic; the
minimum anisotropy of the phonon interaction is $\phi_{AN} \leq
\pi/2 $ for our model interaction of Eq.(3).

\begin{figure}
\noindent
\includegraphics[width=100mm]{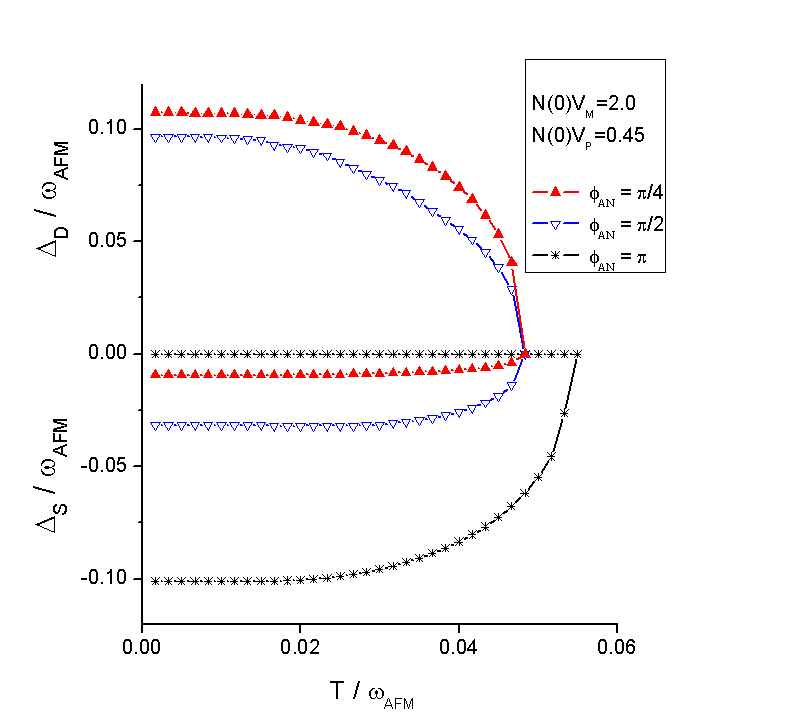}
\caption{(Color online) (D+S) case. Calculated magnitudes of
d-wave components ($\Delta_{D}/\omega_{AFM}$) and s-wave
components ($\Delta_{S}/\omega_{AFM}$) for different anisotropy
angles $\phi_{AN}=\pi/4, \pi/2,$ and $\pi$. For all cases
$N(0)V_{M}=2.0$, $N(0)V_{P}=0.45$ and $\omega_{ph}
/\omega_{AFM}$=0.5. \label{fig1}}
\end{figure}

Fig.1 shows typical results of the (D+S) gap solution.  The
normalized d-wave gap $\Delta_D / \omega_{AFM}$ and the s-wave gap
$\Delta_S /\omega_{AFM} $ are plotted for different values of
anisotropy angle $\phi_{AN}=\pi/4, \pi/2$ and $\pi$ with fixed
values of coupling constants $N(0)V_{M}=2$ and $N(0)V_{P}=0.45$.
The effective dimensionless coupling constants $\lambda_{AFM,ph}$
are smaller than theses values which is defined below as a
projected average with a SC gap function.

\begin{equation}
\lambda_{AFM,ph}= N(0) \frac{\sum_{\phi,\phi^{'}} V_{AFM,ph}(\phi
-\phi^{'}) \eta(\phi) \eta(\phi^{'})}{\sum_{\phi} \eta^2 (\phi)}
\end{equation}
\noindent where $\eta(\phi)=\cos(2\phi)$ for d-wave gap and
$\eta(\phi)=1$ for s-wave gap.
For the chosen potential strengths $N(0)V_{M}=2$ and
$N(0)V_{P}=0.45$ in Fig.1, the effective coupling strengths are
$\lambda_{AFM,D}=0.332$ and $\lambda_{ph,S}=0.450$, 0.338, 0.198
for $\phi_{AN}=\pi, \pi/2$, $\pi/4$, respectively.

When $\phi_{AN}=\pi$ (corresponding to a perfectly isotropic
phonon interaction), solution is either a pure d-wave or a pure
s-wave gap primarily depending on the strength of $\lambda_{AFM}$
and $\lambda_{ph}$; the final dominant instability is determined
not only by the interaction strengths but also with the cutoff
energy scales $\omega_{AFM}$ and $\omega_{ph}$.
In this case, $\lambda_{ph,S}$ (=0.450) is much bigger than $
\lambda_{AFM,D}$ (=0.332), therefore a pure s-wave gap becomes a
solution.
Only when $\phi_{AN} \leq \pi/2$ and $\lambda_{ph,S}$ is not
dominant over $\lambda_{AFM,D}$, a s-wave gap component can
coexist with a d-wave component. When these conditions satisfied,
no matter how weak the phonon interaction is, a finite s-wave
component coexists over the entire temperatures and shares the
same transition temperature T$_c$. Also this T$_{c}$ remains the
same as the pure d-wave T$_{c0}$ without the phonon interaction,
regardless of the presence of $\Delta_{S}$ and its magnitude. This
somewhat unexpected result can be understood from the coupled gap
equations Eq.(7-8). As far as the d-wave component is nonzero,
that is the dominant pairing gap and T$_c$ is determined solely by
the d-wave gap equation Eq.(7) in the limit of $\Delta_{D},
\Delta_{S} \rightarrow 0$.
In the opposite limit, namely, when $\lambda_{ph,S}$ is dominant
over $\lambda_{AFM,D}$, then even if $\phi_{AN} \leq \pi/2$, the
d-wave gap is entirely suppressed and T$_c$ is determined solely
by the s-wave gap equation Eq.(8).

\begin{figure}
\noindent
\includegraphics[width=100mm]{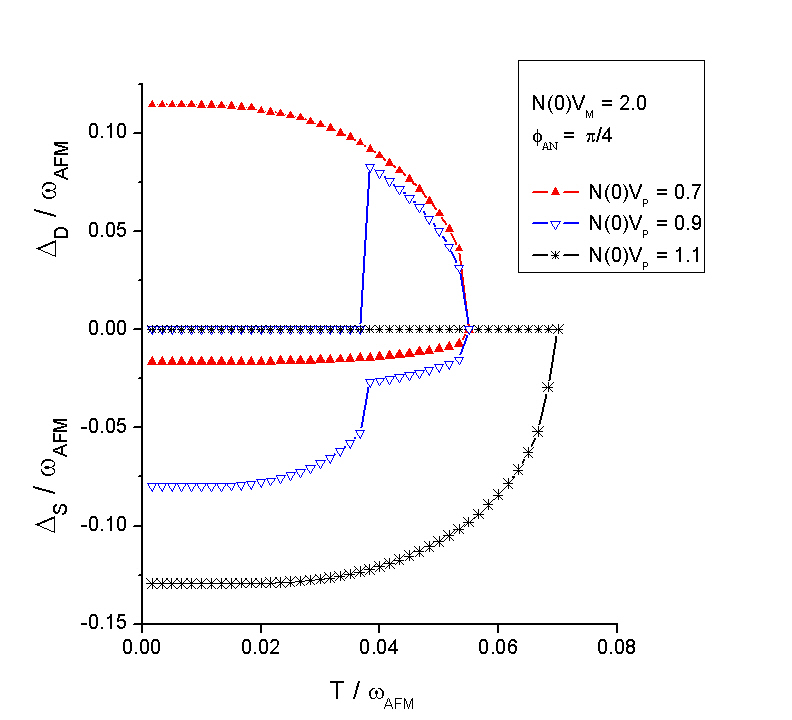}
\caption{(Color online) (D+S) case. Calculated magnitudes of
d-wave components ($\Delta_{D}/\omega_{AFM}$) and s-wave
components ($\Delta_{S}/\omega_{AFM}$) for different strength of
phonon interaction $N(0)V_{P}=0.7, 0.9,$ and $1.1$. For all cases
$N(0)V_{M}=2.0$, $\phi_{AN}=\pi/4$ and $\omega_{ph}
/\omega_{AFM}$=0.5. \label{fig2}}
\end{figure}

In Fig.2 we fixed the anisotropy angle $\phi_{AN} (=\pi/4$) and
changed the coupling strength of phonon $N(0)V_{P}$ (=0.7, 0.9,
1.1). The variation of the magnitude of the s-wave gap
$\Delta_{S}$ can be understood with the effective coupling
strength of $\lambda_{ph,S}$ (=0.31, 0.39, 0.48, respectively)
compared to $\lambda_{AFM,D}$ (=0.332) as in Fig.1. One peculiar
feature occurs when $N(0)V_{P}=0.9$ ($\lambda_{ph,S}$=0.39)(down
triangles). The higher transition temperature T$_{c,high}$ is
understood as the coexistence cases of Fig.1; although
$\lambda_{ph,S}~ (=0.39)
> \lambda_{AFM,D}~ (=0.332)$, the fact $\omega_{AFM} >
\omega_{ph}$ makes the d-wave pairing still slightly dominant at
high temperatures. But at lower temperature one more transition
occurs where the d-wave gap collapses to zero and the s-wave gap
component makes a sudden increase.
This abrupt change of gaps is a first order transition and
indicates that there are two competing local minima in the free
energy. With temperature the global minimum changes  from one
local minimum to the other local minimum. This peculiar behavior
is due to the closeness between $\lambda_{AFM,D}$ and
$\lambda_{ph,S}$, and yet the distance of the energy scale between
$\omega_{AFM}=1$ and $\omega_{P}=0.5$ as we have chosen in our
calculations (all energy scales are normalized by $\omega_{AFM}$
in this paper) in this particular case.
This is a rather artificial result of our model; nevertheless, it
is an interesting behavior of the (D+S) type gap equations of
Eq.(7-8).

The summary for the (D+S) type gap is: (1) The mixed type gap
solution is possible if the phonon interaction has a proper
anisotropy and its coupling strength is subdominant to the d-wave
pairing strength; (2) T$_c$ is not enhanced as far as the d-wave
gap remains finite regardless of the presence and magnitude of the
s-wave gap component.

\subsection {(D+$i$S) case}

\begin{figure}
\noindent
\includegraphics[width=100mm]{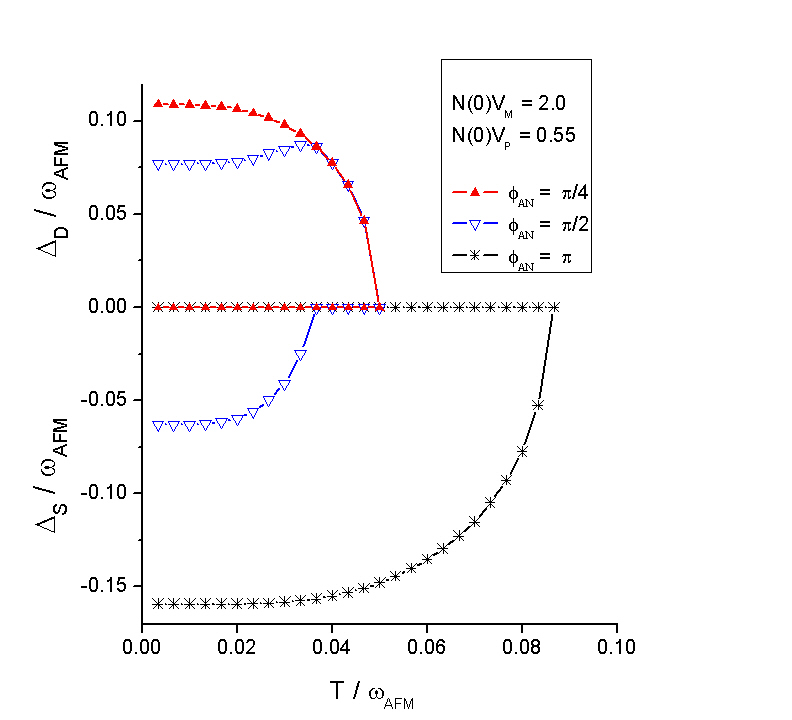}
\caption{(Color online) (D+$i$S) case. Calculated magnitudes of
d-wave components ($\Delta_{D}/\omega_{AFM}$) and s-wave
components ($\Delta_{S}/\omega_{AFM}$) for different anisotropy
angles $\phi_{AN}=\pi/4, \pi/2,$ and $\pi$. For all cases
$N(0)V_{M}=2.0$, $N(0)V_{P}=0.5$ and $\omega_{ph}
/\omega_{AFM}$=0.5. \label{fig3}}
\end{figure}

In this case, we assume $\Delta_{AFM}(k)=\Delta_{d} \cos(2 \phi)$
and $\Delta_{ph}=i \Delta_{s}$. Accordingly, $\Delta_{t}
(\phi)=\Delta_{d} \cos(2 \phi)+ i \Delta_{s}$ and $E(\phi) = \sqrt
{\xi^2 + \Delta_{d}^2(\phi) + \Delta_{s}^2}$. The gap equations
are
\begin{eqnarray}
\Delta_d (\phi) & = & - \sum_{\phi^{'}} V_{AFM}(\phi - \phi^{'})
\Delta_t (\phi^{'})  \chi(\phi^{'},\omega_{AFM}), \\
i \Delta_s (\phi) & = &  -\sum_{\phi^{'}} V_{ph}(\phi - \phi^{'})
\Delta_t (\phi^{'}) \chi(\phi^{'},\omega_{ph})
\end{eqnarray}

The main difference from the (D+S) case is that because of the
phase $i$ between $\Delta_{d}(\phi)$ and $\Delta_{s}$, two gaps
are only indirectly coupled through the quasiparticle energy
$E(\phi)$ in the pair susceptibility $\chi(\phi, \omega_{M,ph})$.
Therefore the coupling of two gaps is much smoother than the (D+S)
case. One consequence of it is that the gap equations can allow
two transition temperatures T$_{c,D}$ and T$_{c,S}$ for each gap
$\Delta_{D}$ and $\Delta_{S}$, respectively. This feature is shown
in Fig.3 where $N(0)V_M =2.0$ ($\lambda_{AFM,D}=0.332$) and
$N(0)V_P =0.55$ are fixed and the anisotropy angle of phonon
interaction $\phi_{AN}$ is varied as $\pi/4$, $\pi/2$, and $\pi$
(correspondingly $\lambda_{ph,S}$ = 0.243, 0.414 and 0.55). Main
features are summarized as: (1) when $\lambda_{ph,S}$ is too weak,
the effect of phonon interaction is totally ignored and the gap
and T$_c$ is solely determined by the AFM interaction (see
$\phi_{AN}=\pi/4$ case). (2) when $\lambda_{ph,S}$ is much
stronger than $\lambda_{AFM,D}$, the d-wave gap is totally
suppressed and the gap and T$_c$ is determined solely by phonon
interaction (see $\phi_{AN}=\pi$ case). This behavior is similar
to the (D+S) case. (3) when the effective coupling strengthes are
comparable, namely $\lambda_{AFM,D} \approx \lambda_{ph,S}$, both
d-wave and s-wave gaps have separate transition temperatures
T$_{c,D}$ and T$_{c,S}$ and two gaps can coexist at low
temperatures (see $\phi_{AN}=\pi/2$ case). In this case, T$_{c,D}$
is always higher transition temperature and this transition
temperature is just the same as the pure d-wave T$_{c0}$ without
phonon interaction. At the lower transition temperature T$_{c,S}$,
the s-wave gap $i \Delta_S$ appears and the magnitude of d-wave
gap $\Delta_D$ accordingly decreases.

To clarify the roles of the anisotropy and the strength of phonon
interaction, in Fig.4 we fixed anisotropy angle as
$\phi_{AN}=\pi$, which simulates a perfectly isotropic phonon
interaction, and varied the interaction strength. Fig.4 shows the
results of these calculations. With $N(0) V_M$ =2.0
($\lambda_{AFM}$=0.332), we varied $N(0) V_P$ (= 0.3, 0.4, 0.5;
accordingly $\lambda_{s}$=0.3, 0.4, 0.5). This plot demonstrates
that the anisotropy of phonon interaction doesn't play a
particular role in the case of the (D+$i$S) type gap. The mixed
gap solution of the (D+ $i$S) type is still possible when
$\lambda_{AFM,D}$ and $\lambda_{ph,S}$ are of the comparable
strength regardless of the anisotropy of phonon interaction.

The summary for the (D+$i$S) type gap is: (1) The mixed type gap
solution is possible if the phonon interaction $\lambda_{ph,S}$ is
comparable but still subdominant to the AFM interaction strength
$\lambda_{AFM,D}$. (2) In contrast to the (D+S) case, separate two
transition temperatures T$_{c,D}$ and T$_{c,S}$ exist for each gap
$\Delta_{D}$ and $\Delta_{S}$; T$_{c,D}$ is always higher than
T$_{c,S}$ and only below T$_{c,S}$ two gaps coexist. (3) As in the
(D+S) case, T$_c$ is not enhanced as far as the d-wave gap remains
finite regardless of the presence and magnitude of the s-wave gap
component. (4) Anisotropy of the phonon interaction is not
particularly necessary.

\begin{figure}
\noindent
\includegraphics[width=100mm]{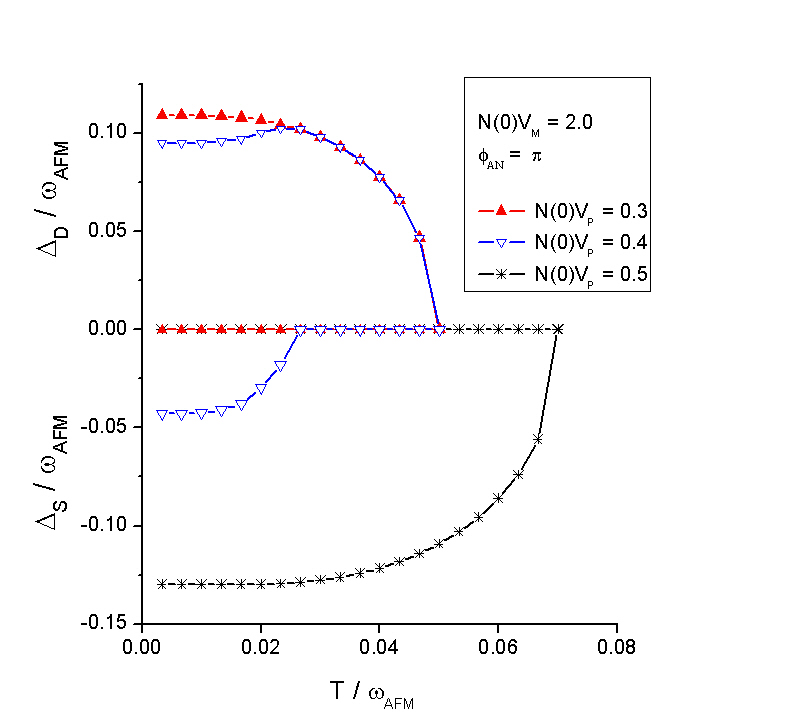}
\caption{(Color online) (D+$i$S) case. Calculated magnitudes of
d-wave component ($\Delta_{d}/\omega_{AFM}$) and s-wave component
($\Delta_{s}/\omega_{AFM}$) for different strength of phonon
interaction $N(0)V_{P}=0.3, 0.4,$ and $0.5$. For all cases
$N(0)V_{M}=2.0$, $\phi_{AN}=\pi$, $\omega_{ph} /\omega_{AFM}$=0.5.
\label{fig4}}
\end{figure}

\begin{figure}
\noindent
\includegraphics[width=100mm]{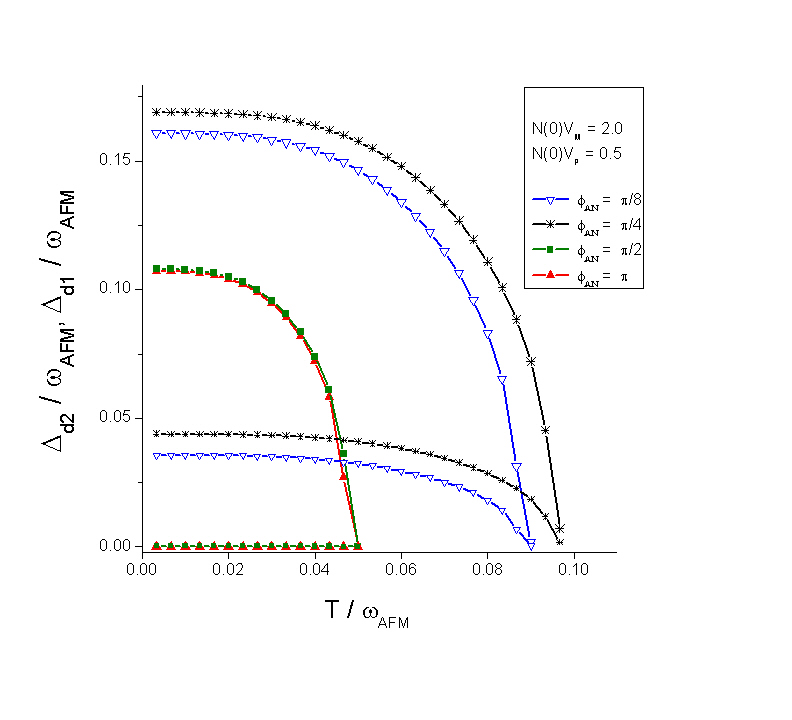}
\caption{(Color online) (D$_{AFM}$+D$_{ph}$) case. Calculated
magnitudes of D$_{AFM}$ component ($\Delta_{d1}/\omega_{AFM}$) and
D$_{ph}$ component ($\Delta_{d2}/\omega_{AFM}$) for different
anisotropy angles $\phi_{AN}=\pi/8, \pi/4, \pi/2,$ and $\pi$. For
all cases $N(0)V_{M}=2.0$, $N(0)V_{P}=0.5$ and $\omega_{ph}
/\omega_{AFM}$=0.5.  \label{fig5}}
\end{figure}

\begin{figure}
\noindent
\includegraphics[width=100mm]{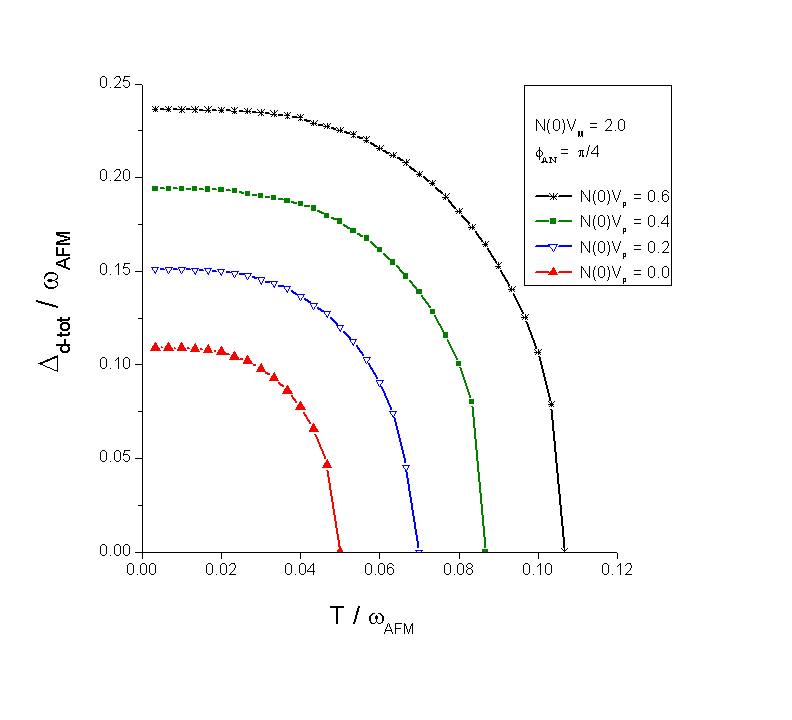}
\caption{(Color online) (D$_{AFM}$+D$_{ph}$) case. Calculated
magnitudes of the total d-wave gap ($\Delta_{d-tot}/
\omega_{AFM}$) for different strength of phonon interaction
$N(0)V_{P}=0.0, 0.2, 0.4,$ and 0.6. For all cases $N(0)V_{M}=2.0$,
$\phi_{AN}=\pi/4$ and $\omega_{ph} /\omega_{AFM}$=0.5.
\label{fig6}}
\end{figure}

\subsection {(D$_{AFM}$+D$_{ph}$) case}

This is the case that phonon interaction also supports the d-wave
pairing. Because it is clear that a perfectly isotropic phonon
interaction has null effect on the d-wave gap, anisotropy of the
phonon interaction is crucial for this case. The total gap is,
therefore, $\Delta_{t} (\phi)=(\Delta_{d1} + \Delta_{d2})\cos(2
\phi)$ and the gap equations are written as

\begin{eqnarray}
\Delta_{d1} (\phi) & = & - \sum_{\phi^{'}} V_{AFM}(\phi -
\phi^{'})
\Delta_t (\phi^{'})  \chi(\phi^{'},\omega_{AFM}), \\
\Delta_{d2} (\phi) & = &  -\sum_{\phi^{'}} V_{ph}(\phi - \phi^{'})
\Delta_t (\phi^{'}) \chi(\phi^{'},\omega_{ph}).
\end{eqnarray}

The above two gap equations can be combined to a single gap
equation but we keep the separate form in order to trace the
effect of the phonon interaction. In contrast to the previous
cases, {\it T$_c$ can be dramatically enhanced}  when the phonon
interaction $V_{ph}(\Delta \phi)$ has a proper degree of
anisotropy. Other authors \cite{honer} also obtained a similar
result using different approaches.
In Fig.5, we plot $\Delta_{d1}$ and $\Delta_{d2}$ separately;
between the same symbols the smaller value is the phonon induced
d-wave gap $\Delta_{d2}$ and the larger one is the AFM induced
d-wave gap $\Delta_{d1}$. We change the anisotropy angle of
$V_{ph}(\Delta \phi)$ with the fixed interaction strengths of
$N(0)V_M=2.0 ~(\lambda_{AFM,D}=0.332$) and $N(0)V_p=0.5$. For the
anisotropy angle $\phi_{AN} = \pi/2$ and $\pi$, the phonon
interaction has absolutely no effect on the d-wave pairing; this
is simply because the d-wave projected average Eq.(10) becomes
zero for these two commensurate angles with our model potential
Eq.(3).  With a stronger anisotropy ($\phi_{AN} = \pi/4$;
$\lambda_{ph,D}=0.12$), the phonon scattering sees only a limited
part of the d-wave gap, therefore the d-wave gap can be seen
effectively as a s-wave gap for the anisotropic phonon. For a even
stronger anisotropy ($\phi_{AN} = \pi/8; \lambda_{ph,D}=0.098$),
too narrow scattering angle reduces the effective coupling
strength. In our model potential, $\phi_{AN} = \pi/4$ is the
optimal anisotropy.
In fact, even weaker anisotropy of $\pi/4 < \phi_{AN} < \pi$ --
except the special commensurate angles $\phi_{AN} = \pi/2$ and
$\pi$ -- produces some uncompensated effective interaction
$\lambda_{ph,D}$ and boosts the d-wave pairing. This is an
artifact of the model potential Eq.(10) which always emphasizes
the small angle scattering. Therefore, for some real phonons, it
is possible to have a repulsive (destructive for d-wave pairing)
$\lambda_{ph,D}$ if large angle (around $\Delta \phi =\pi/2$)
scattering overweighs small angle scattering.

Fig.6 shows the results of total gap $\Delta_{d-tot} =
\Delta_{d1}+\Delta_{d2}$ with the optimal anisotropy angle
$\phi_{AN}=\pi/4$ and varying phonon coupling strength
$N(0)V_P$=0.0, 0.2, 0.4, and 0.6 ($\lambda_{ph,D} =0.0, 0.047,
0.095,$ and 0.142, respectively). The results demonstrate that the
relatively weak phonon interaction $\lambda_{ph,D}$ can
significantly increase T$_c$ from the pure AFM interaction induced
T$_{c0}$.

\begin{figure}
\noindent
\includegraphics[width=100mm]{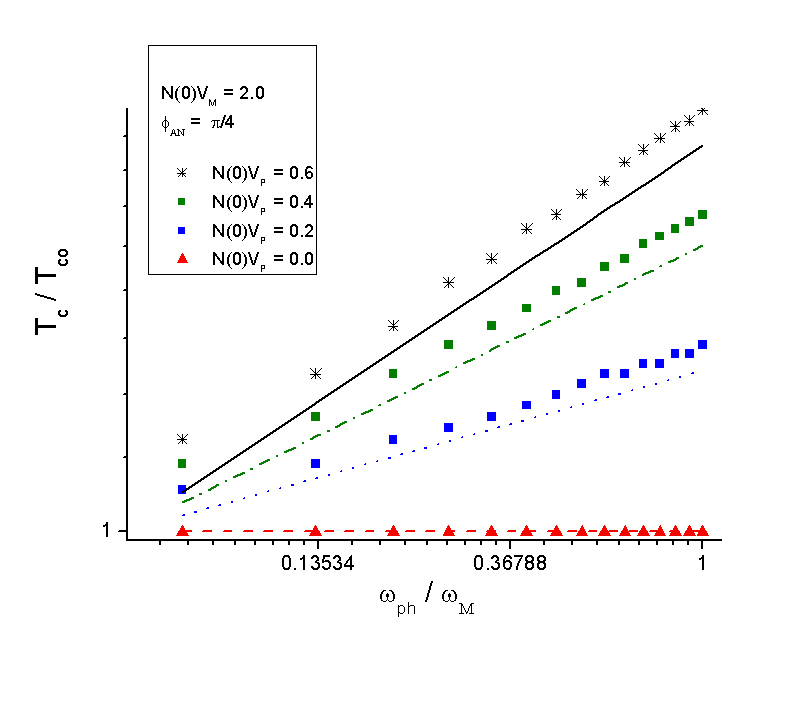}
\caption{(Color online) (D$_{AFM}$+D$_{ph}$) gap. T$_c$ (symbols)
calculated as a function of $\omega_{ph} /\omega_{AFM}$ for
different strength of phonon interaction $N(0)V_{P}=0.0, 0.2,
0.4$, and 0.6, normalized by T$_{c0}$ the transition temperature
with $N(0)V_{P}=0.0$. For all cases, $N(0)V_{M}=2.0$ and
$\phi_{AN}=\pi/4$. Lines are the results of the analytic formula
of T$_c$ Eq.(15). The symbols and lines of the same colors have
the same parameters set. \label{fig7}}
\end{figure}

Considering the experimental findings about phonons in the
high-T$_c$ cuprates, the best candidate for the anisotropic
pairing phonon is B$_{1g}$ buckling mode of the plane oxygen
motion\cite{cuk 04}. Devereaux and coworkers extensively analyzed
the behaviors of B$_{1g}$ mode and found : (1) B$_{1g}$ mode has a
strong anisotropic coupling matrix element concentrating around
antinodal points; (2) the maximum coupling strength can reach as
high as $\lambda \sim 3$.
Compared to our model calculations, the optimal anisotropy for the
d-wave pairing $\phi_{AN} = \pi/4$ is just about the same degree
of anisotropy of the B$_{1g}$ mode. Regarding the coupling
strength, for example, in Fig.6 the maximum coupling strength
$\lambda_{max}$ is $N(0) V_p$ =0.6 and the corresponding average
strength is $\lambda_{ph,D}$= 0.142. These values are much smaller
than the ones extracted by Devereaux and coworkers\cite{cuk 04}.
Larger values of coupling constants would be more realistic.
However since our study is limited within the BCS gap equations we
chose smaller values of coupling constants. With larger coupling
strength ($\lambda > O(1)$), we should go for the strong coupling
theory for the reliable analysis. Also for the same phonon, the
coupling strength extracted from ARPES experiments and the
coupling strength for the pairing problem can be very different
due to the on-site Coulomb interaction.

Another important anisotropic phonon in Bi-2212 is the breathing
mode which strongly interacts near the nodes. The effect of this
phonon is partially included in our calculations because our model
phonon interaction Eq.(3) only constraints the scattering angle by
$\phi_{AN}$ but doesn't constraint the incoming {\bf k} and
outgoing momenta  {\bf k$^{'}$} ($\phi$ and $\phi^{'}$ in our
model). The anisotropic phonon scattering concentrating near nodes
will be mostly averaged out by the sign changing d-wave gap and
will be of no importance for the d-wave pairing. For realistic
estimate of the phonon effects in HTC compounds, all important
phonons should be included to calculate $\lambda_{ph,D}$ using
full coupling matrix elements. This is beyond the scope of the
current paper.

Since we found that an anisotropic phonon can dramatically enhance
T$_c$ of d-wave pairing, we have to consider the isotope effect of
the phonon, which is reported to be anomalously small, in
particular, near optimal doping region by many
experiments\cite{isotope}. For this purpose, Fig.7 shows the
calculated T$_c$'s (symbols) from Eq.(13-14) as a function of the
phonon energy cutoff $\omega_{ph}/\omega_{AFM}$ for various phonon
coupling strength $N(0)V_{P} = 0.0, 0.2, 0.4$ and 0.6,
respectively. The BCS theory predicts T$_c \sim \omega_{ph}$
($\alpha=0.5$), which should show up as a linear lines in Fig.7.
Our numerical results show much weaker power than the linear one
indicating $\alpha < 0.5$. For more analytic investigation for the
isotope effect and the origin of the T$_c$ enhancement by phonon
in the (D$_{AFM}$+D$_{ph}$) case, we derived an analytic T$_c$
equation. The gap equations Eq.(13-14) can reduce to a single
T$_c$ equation by adding two equations and taking a limit of
$\Delta_{d1}, \Delta_{d2} \rightarrow 0$. The pair susceptibility
$\chi(\omega_{AFM,ph})$ (Eq.(9)) is integrated out using BCS
approximation which is $\int^{\omega_{D}} _{0} d\xi [\tanh
\frac{\xi}{2 T}] / \xi \approx \ln [2 C ~\omega_{D} / \pi T] $ (C
= $e^{\gamma} \approx 1.7807$), valid when $\omega_{D}/2 T_c \gg
O(1)$. Finally we obtained the T$_c$ formula of the
(D$_{AFM}$+D$_{ph}$) case as

\begin{equation}
T_c \simeq 1.13~ \omega_{AFM} ^{\tilde{\lambda}_{AFM}} \cdot
\omega_{ph} ^{\tilde{\lambda}_{ph}} ~~e^{-1/\lambda_t} .
\end{equation}
where $\lambda_t =(\lambda_{AFM} +\lambda_{ph} )$,
$\tilde{\lambda}_{AFM} =\lambda_{AFM} /\lambda_t $ and
$\tilde{\lambda}_{ph} =\lambda_{ph} /\lambda_t $, and
$\lambda_{AFM}$ and $\lambda_{ph}$ are the dimensionless effective
coupling constants obtained from Eq.(10) with d-wave gap average
for both couplings.
$\omega_{AFM}$ and $\omega_{ph}$ are the energy cutoffs of the AFM
interaction and phonon interaction, respectively. As mentioned
above, this T$_c$ formula is derived assuming $\omega_{M,ph}/2 T_c
\gg O(1)$ and this condition is well satisfied when $\lambda_t <
O(1)$. In Fig.7, we also plot the results of T$_c$'s from Eq.(15)
(lines) with the same parameters and compare them with the
numerical calculations (symbols) from Eq.(13-14). It shows a
reasonably good agreement between two results although the
deviations increase with increasing the coupling constant $N(0)
V_p$ (or $\lambda_{ph,D}$) as expected. However, the overall and
quantitative behavior of T$_c$ is well captured by Eq.(15).

Now we are in position to read the phonon isotope coefficient
$\alpha$ from Eq.(15), which is

\begin{equation}
\alpha = \frac{1}{2}\tilde{\lambda}_{ph} =\frac{1}{2}
\frac{\lambda_{ph}}{\lambda_{AFM} + \lambda_{ph}}.
\end{equation}

For example, with representative values of $\lambda_{AFM}=0.33$,
$\lambda_{ph}=0.1$ and $\omega_{ph} /\omega_{AFM} =0.5$, we obtain
$\alpha \approx 0.116$, which is pretty small value compared to
the standard BCS value of 0.5 while T$_{c}$ is enhanced by about
100$\%$ from T$_{c0}$ the transition temperature without phonon
interaction (see Fig.7; $N(0)V_P =0.6$ is $\lambda_{ph}=0.142$).
This isotope coefficient equation Eq.(16) provides a very
plausible resolution why the isotope coefficient is so low near
optimal doping region where T$_c$ is the highest and increases
toward underdoped regime (decreasing T$_c$). The phonon coupling
strength ($\lambda_{ph}$) is likely unchanged with doping but the
effective coupling strength of the AFM mediated interaction
($\lambda_{AFM}$) will change sensitively with doping. If we
assume that $\lambda_{AFM}$ increases with increasing doping but
$\lambda_{ph}$ remains constant, Eq.(16) describes the general
trend of the experimentally observed oxygen isotope effect
$\alpha_{O}$ \cite{isotope}.

\begin{figure}
\noindent
\includegraphics[width=100mm]{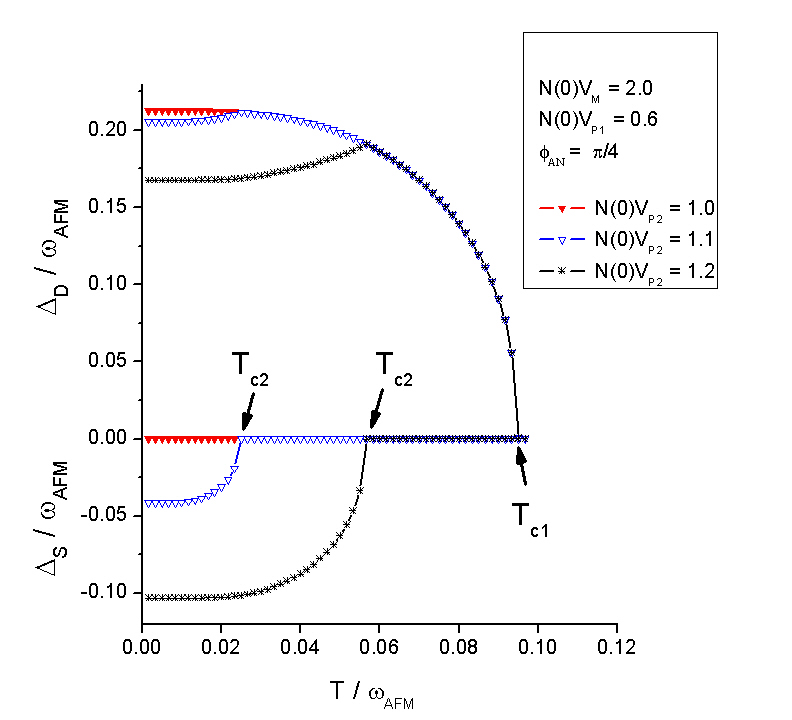}
\caption{(Color online) (D$_{AFM}$+D$_{ph}$ +$i$S$_{ph}$) case.
Calculated magnitudes of total d-wave component ($\Delta_D =
\Delta_{AFM}+\Delta_{ph}$) and s-wave component ($\Delta_{s}$) for
different strength of phonon interaction $N(0)V_{P2}=1.0, 1.1$ and
$1.2$. For all cases $N(0)V_{M}=2.0$, $N(0)V_{P1}=0.6$,
$\phi_{AN}=\pi/4$, and  $\omega_{p} /\omega_{AFM}$=0.5.
\label{fig8}}
\end{figure}

\subsection {(D$_{AFM}$+D$_{ph}$ +$i$S$_{ph}$) case}

Combining the (D$_{AFM}$+D$_{ph}$) and (D+ $i$S) case, we can
consider the (D$_{AFM}$+D$_{ph}$ +$i$S$_{ph}$) type solution. This
is indeed a natural solution of the combined gap equations (12),
(13), and (14) and the total gap  will be $\Delta_{t}
(\phi)=(\Delta_{d1} + \Delta_{d2})\cos(2 \phi) + i \Delta_{s}$. In
this case, phonon interaction(s) mediate both d-wave and s-wave
pairings, which seems contradicting to a common knowledge. The
fact is that phonon alone - no matter how anisotropic - cannot
induce a d-wave gap but it can boost it if a d-wave gap is formed
by other interaction. Then as we found in the (D+ $i$S) section, a
phonon interaction can add $i$S component at lower temperatures.
The interesting point is that the phonon interaction boosting
d-wave gap and the phonon interaction inducing $i$S gap can
originate from the same phonon if the anisotropy and the
interaction strength are properly tuned. Of course more general
case is that two different phonon modes play separate roles,
respectively. For clearness, we rewrite the gap equations
describing (D$_{AFM}$+D$_{ph}$ +$i$ S$_{ph}$) case below.

\begin{eqnarray}
\Delta_{d1} (\phi) & = & - \sum_{\phi^{'}} V_{AFM}(\phi -
\phi^{'})
\Delta_t (\phi^{'})  \chi(\phi^{'},\omega_{AFM}), \\
\Delta_{d2} (\phi) & = &  -\sum_{\phi^{'}} V_{1,ph}(\phi -
\phi^{'})
\Delta_t (\phi^{'}) \chi(\phi^{'},\omega_{p1}), \\
i \Delta_s (\phi) & = &  -\sum_{\phi^{'}} V_{2,ph}(\phi -
\phi^{'}) \Delta_t (\phi^{'}) \chi(\phi^{'},\omega_{p2})
\end{eqnarray}

\noindent where the pair susceptibilities $\chi (\phi, \omega)$
are defined in Eq.(9) and the quasiparticle energy is $E(\phi) =
\sqrt {\xi^2 + \Delta_{d}^2(\phi) +\Delta_{s}^2}$ with $\Delta_{d}
(\phi)=(\Delta_{d1} + \Delta_{d2})\cos(2 \phi)$. For generality we
assume two different phonon interactions  $V_{1,ph}$ and
$V_{2,ph}$, respectively but they can be the same phonon as
explained above.
\begin{figure}
\noindent
\includegraphics[width=100mm]{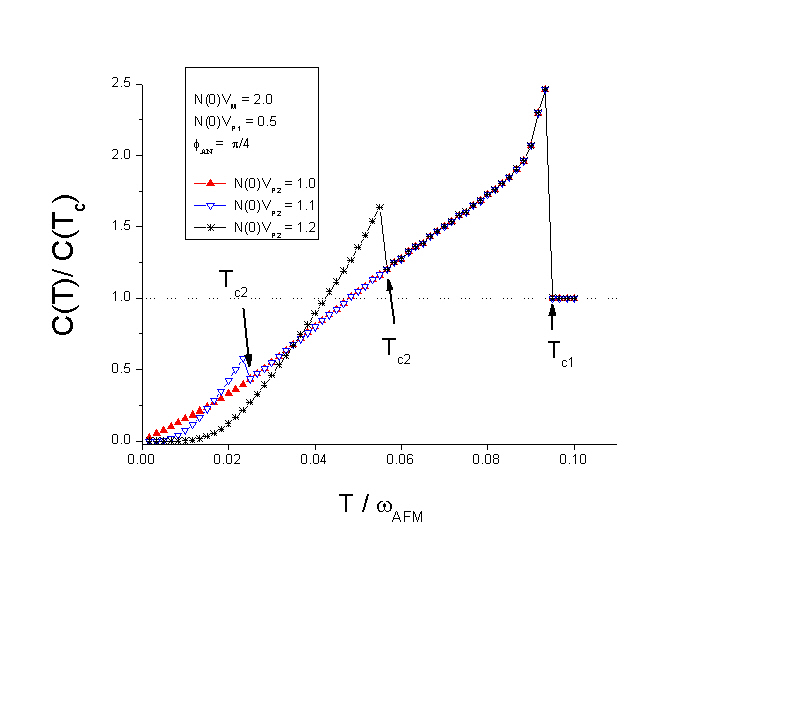}
\vspace{-2.5cm} \caption{(Color online) (D$_{AFM}$+D$_{ph}$ +$i$
S$_{ph}$) case. Normalized specific heats C(T)/C(T$_c$) for
different strength of phonon interaction $N(0)V_{P2}$. There are
two transition temperatures T$_{c1}$ (higher) and  T$_{c2}$
(lower). Same parameters as in Fig.8. \label{fig9}}
\end{figure}

In Fig.8 we plot the solutions of
$\Delta_{d}=\Delta_{d1}+\Delta_{d2}$ and $\Delta_{s}$ components
separately. We fix the AFM interaction $N(0)V_M$=2.0 and the
d-wave boosting phonon interaction $N(0)V_{P1}$=0.6 as a case of
Fig.6 and vary the phonon interaction $N(0)V_{P2}$ inducing $i$S
gap component as 1.0, 1.1, and 1.2, respectively. T$_c$ is
enhanced by $N(0)V_{P1}$ and doesn't change for different
$N(0)V_{P2}$ values because it is determined only by the d-wave
gap component. We also chose the same anisotropy $\phi_{AN}=\pi/4$
for both phonons $V_{1,ph}$ and $V_{2,ph}$ for convenience.
Anisotropy condition for the d-wave boosting phonon is important;
optimal anisotropy is $\phi_{AN}= \pi/4$ for our model. However,
the phonon interaction $V_{2,ph}$ inducing $i$S gap doesn't need
to be anisotropic.

This type of gap solution can provide a possible resolution to the
recent experiments of HTC which indicate the presence of a s-wave
component at low temperatures. The anomalous tunnelling
conductance in YBCO\cite{tunnelling}, non-monotonic behavior of
the penetration depth at low temperatures and its isotope
effect\cite{penet}, etc indicate a mixed gap of the (D+$i$S) type
at low temperatures and a phonon effect with it. Fig.9 show the
normalized specific heat $C(T)/C(T_{c1})$ for the gap solutions of
Fig.8. General feature is that it shows a d-wave type specific
heat below T$_{c1}$ and then turns into a s-wave type behavior
below T$_{c2}$ where $i$S gap component opens. Since the second
transition at T$_{c2}$ is a true second order phase transition,
the specific heat exhibits a jump. But the size of jump is far
smaller than the BCS value of $\Delta C=1.43 ~C(T_{c2})$. The
reason is because when $i$S gap opens at T$_{c2}$, the magnitude
of d-wave gap also abruptly decreases so that the amount of the
specific heat jump is partially cancelled. Nevertheless, careful
experiments should be able to discern the jump in the specific
heat at the second transition temperature T$_{c2}$ if this
scenario is true.

\begin{figure}
\noindent
\includegraphics[width=100mm]{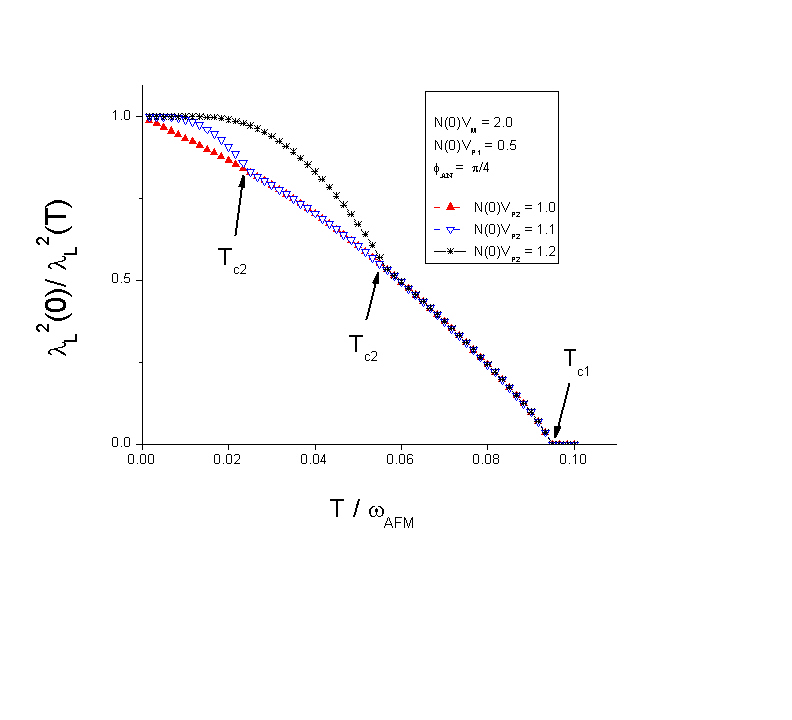} \vspace{-2.5cm}
\caption{(Color online) (D$_{AFM}$+D$_{ph}$ +$i$ S$_{ph}$) case.
Normalized superfluidity density $\lambda_{L} ^2$(0)/$\lambda_{L}
^2$(T) for different strength of phonon interaction $N(0)V_{P2}$.
Same parameters as in Fig.8. \label{fig10}}
\end{figure}

Fig.10 shows the normalized superfluid density $\lambda_{L} ^2
(0)/\lambda_{L} ^2 (T)$ for the gap solutions of Fig.8. The
general features are the same as in the specific heat: d-wave
behavior in between T$_{c1}$ and T$_{c2}$, and s-wave behavior
below T$_{c2}$. In particular, the additional sharp increase of
$\lambda_{L} ^{-2} (T)$ below T$_{c2}$ has some resemblance to the
experimental observations \cite{penet} although the flattened out
behavior at very low temperatures is not clearly observed
experimentally.

We propose a firm experimental test for this scenario. The d-wave
boosting phonon interaction does not have a minimum interaction
strength; as far as it has a proper anisotropy it will enhance the
d-wave pairing as much as with its strength. For the $i$S
component to appear below T$_{c2}$ the phonon interaction needs a
minimum strength to overcome a part of d-wave component.
Therefore, a controlled suppression of d-wave gap (with
non-magnetic impurities not to weaken the s-wave pairing) will
widen a window of low temperature region where the $i$ S component
pops up.

\section{Conclusions}

In this paper we studied the effects of phonon interaction on the
superconducting pairing in the background of d-wave gap already
formed by the AFM interaction. In particular, we studied the role
of anisotropy of the phonon interaction and possible multigap type
solutions within a generalized BCS theory. For many cases the
anisotropy of the phonon interaction is a crucial condition for
interesting interplay with the AFM interaction but not always; the
(D+S) type gap need anisotropic phonon but (D+$i$S) type gap is
more tolerable with the anisotropy of phonon interaction because
two gaps are more gently coupled in the (D+$i$S) case. In both
cases, T$_c$ is not enhanced at all by phonon interaction as far
as the d-wave gap component remains finite.

Anisotropic phonon can boost T$_c$ together with the AFM
interaction in the ($D_{AFM}$+$D_{ph}$) type solution. With
numerical calculations and analytic T$_c$ equation, we showed that
T$_c$ is enhanced dramatically following the modified BCS
exponential form T$_c$ $\sim \exp{( -1/\lambda_t)}$ with the total
coupling strength $\lambda_t = \lambda_{AFM} + \lambda_{ph}$. This
type of solution can also explain how and why the phonon isotope
effect is strongly reduced despite the large enhancement of T$_c$
by phonon. Our isotope coefficient formula Eq.(16) provides a good
description of the overall trend of oxygen isotope coefficient
$\alpha_O$ \cite{isotope}.

Also a combined type solution of  ($D_{AFM}+D_{ph}+i$S) gap is
considered. This type of gap not only shows the features of the
$(D_{AFM}+D_{ph})$ gap -- enhanced T$_c$ and small isotope effect
-- but also shows the appearance of a s-wave component at low
temperatures. The latter feature can provide a microscopic origin
of the small s-wave component suggested from recent experiments of
tunnelling conductance \cite{tunnelling} and penetration depth
measurements \cite{penet}. Calculations of the superfluidity
density showed a qualitative agreement with experiments of the
abrupt increase of $\lambda^{-2} _L (T)$ at lower temperatures
\cite{penet}. However, this scenario should be confirmed by
experimental observation of a specific heat jump at the second low
transition temperature T$_{c2}$ for the s-wave gap component as
shown in Fig.9. We proposed a systematic impurity doping study to
test this scenario.

Finally, we would like to restate the limitation of our work.
First, the dynamical interplay between phonon(s) and AFM
interaction is not properly treated. This problem was already
studied by many authors\cite{polaron,honer,Zeyher}. We bypassed
this complicated problem on the phenomenological basis; after all
correlation effects taken place we assumed that fermion
quasiparticles remain, and AFM spin fluctuations and several
phonon modes are considered as experimentally defined quantities
after mutual screening and renormalization. Second, when we
consider the s-wave pairing, we should have included the on-site
Coulomb interaction in the HTC cuprates in addition to the phonon
interaction. Therefore, $\lambda_{ph,S}$ in this paper should be
considered as a renormalized quantity after including the on-site
Coulomb interaction. Perhaps, this is the reason why our
$\lambda_{ph,S}$ ($\sim 0.1$) is order of magnitude smaller than
the experimental values\cite{cuk 04} $\lambda_{exp} \sim O(1)$.

I thank J. Zaanen for drawing my attention to this problem and
discussions during MSM07 in Khiva. Y. B. was supported by the
KOSEF through the CSCMR and the Grant No. KRF-2007-070-C00044,
KRF-2007-521-C00081.

\end{document}